\documentclass[12pt,preprint]{aastex}

\usepackage{amsmath}

\begin{document}

\title{Formation of Millisecond Pulsars with Low-Mass Helium White Dwarf Companions
in Very Compact Binaries}

\author{Kun Jia$^{1}$ and X.-D. Li$^{1,2}$}

\affil{$^{1}$Department of Astronomy, Nanjing University, Nanjing 210093, China}

\affil{$^{2}$Key laboratory of Modern Astronomy and Astrophysics (Nanjing University), Ministry of
Education, Nanjing 210093, China}

\affil{$^{}$lixd@nju.edu.cn}

\begin{abstract}

Binary millisecond pulsars (BMSPs) are thought to have evolved from low-mass X-ray
binaries (LMXBs). If the mass transfer in LMXBs is driven by nuclear evolution of the
donor star, the final orbital period is predicted to be well correlated with the mass
of the white dwarf (WD), which is the degenerate He core of the donor.
Here we show that this relation
can be extended to very small WD mass ($\sim 0.14-0.17\,M_{\sun}$) and narrow orbital
period (about a few hours), mainly depending on the metallicities of the donor stars.
There is also discontinuity in the relation, which is due to the temporary
contraction of the donor when the H-burning shell crosses the hydrogen discontinuity.
BMSPs with low-mass He WD companions in very compact binaries
can be accounted for if the progenitor binary
experienced very late Case A  mass transfer. The WD companion of PSR J1738+0333 is
likely to evolve from a Pop~II star. For PSR J0348+0432, to explain its extreme compact
orbit in the Roche lobe-decoupling phase, even lower metallicity ($Z=0.0001$) is required.

\end{abstract}


\section{Introduction}
Millisecond pulsars (MSPs) are old neutron stars (NSs) with  short
spin period (${P}_\mathrm{s}\lesssim 30$ ms) and weak surface
magnetic fields ($B\sim 10^{8}-10^{9}$ G)\citep{Lorimer2008}. Soon
after the discovery of the first MSP B1937+21  \citep{Backer1982}, a
scenario was proposed to link the evolution of low-mass X-ray
binaries (LMXBs) with binary millisecond pulsars (BMSPs)
\citep{Alpar1982,Radhakrishnan1982}. A NS in LMXBs may accrete mass
and angular momentum from its companion star for sufficiently long
time (usually a few Gyr), and subsequently be spun up to millisecond
spin periods\footnote{ A large fraction of LMXBs may have evolved
from binaries with an intermediate-mass secondary
\citep[e.g.,][]{Podsiadlowski2002}.  However, most of the spin-up
processes should occur during the LMXB phase.}. After the mass
transfer terminates, the radio radiation from the NS starts to turn
on, leaving a MSP accompanied by a low-mass ($\lesssim{0.4\
M_{\odot}}$) white dwarf (WD) \citep[][for
reviews]{Bhattacharya1991,Tauris2006}. This recycling scenario has
been strongly supported by the discovery of several X-ray pulsars
with millisecond periods in LMXBs \citep[see][for a
summary]{Patruno2012}, and the transition between a LMXB and a MSP
for PSR J1023+ 0038 \citep{Archibald2009} and IGR J18245$-$2452
\citep{Papitto2013}.

The evolution of LMXBs depends on the initial orbital periods
($P_\mathrm{i}$) when Roche-lobe overflow (RLO) occurs, and there
exists a so-called bifurcation period
\citep{Pylyser1988,Pylyser1989}. Its value is around 1 day and
relies heavily on the mechanism and efficiency of the orbital
angular momentum loss (AML) \citep{Ergma1998,Podsiadlowski2002,van
der Sluys2005b,Ma2009}. If $P_\mathrm{i}$ is below the bifurcation
period, the donor does not evolve much at the beginning of RLO, and
the orbit continues to contract, along the cataclysmic varible-like
or ultra-compact X-ray binary (UCXB) evolutionary tracks
\citep{Deloye2008,Lin2011}. When $P_\mathrm{i}$ is above the
bifurcation period, the mass transfer will begin with an evolved,
red giant  donor. Due to the nuclear evolution of the donor star,
the orbit expands until the donor star exhausts its nuclear fuel,
forming a low-mass WD. For low-mass ($<2.3\ M_{\odot}$) red giant
stars there exists a unique relationship between the degenerate He
core mass and the radius of the star
\citep{Refsdal1971,Webbink1983}, thus resulting in a relation
between the WD mass $M_\mathrm{WD}$ and the final orbital period
$P_\mathrm{f}$ at the end of mass transfer
\citep{Joss1987,Rappaport1995,Tauris1999,DeVito2010}.

Although this recycling scenario is now widely accepted for MSPs,
some aspects such as the details of accretion process are still not
clear \citep{Tauris2006}. For example, according to statistical
analysis of the pulsar masses, the mean mass of MSPs is about
$0.2~M_{\odot}$ higher than that of non-recycled PSRs
\citep{Zhang2011}, which implies that the mass transfer is
non-conservative in general, i.e. only a small fraction of the
transferred matter is accreted and the rest has escaped from the
systems.

Previous works mainly focus on the $P_\mathrm{f}-M_\mathrm{WD}$ relation for wide systems
with $P_\mathrm{f}>1$ day, and pay less attention to compact MSP/He WD systems.
In recent years, several compact MSP/He WD binary systems were discovered, including
PSRs J0348+0432 \citep{Antoniadis2013}, J0751+1807 \citep{Nice2008}, J1012+5307
\citep{van Kerkwijk2005}, J1738+0333 \citep{Antoniadis2012}, and J1910$-$5959A \citep{Corongiu2012}. These pulsars all have short ($<1$ day) orbital periods and
low-mass ($<0.2\,M_{\sun}$) WDs. In particular, the
extremely short orbital period (2.46 hr) of PSR J0348+0432 is difficult to explain in
the traditional evolutionary theory \citep{Antoniadis2013}.


The aim of this work is to investigate the formation of compact
($\lesssim1$ day) MSP/He WD systems, and obtain a complete
$P_\mathrm{f}-M_\mathrm{WD}$ relation down to the short-period end.
A similar work was recently done by \cite{Smedley2014}, and the
authors were concentrating on BMSPs in the Galactic field. Here we
extend the relation to lower metallicities, and apply it to the
formation of PSRs J0348+0432 and J1738+0333.

The rest of this paper is organized as follows. We first describe the binary evolution code
used in this work in Section 2. We present the calculated results for a series of
binary models in Section 3, and compare them with the observations of two BMSPs in Section 4.
We summarize the results in Section 5.

\section{Binary evolution code}
We adopt an updated stellar evolution code EV originally developed by
\cite{Eggleton1971,Eggleton1972}\cite[also see][]{Han2004,Pols1995} to calculate the evolution \textbf{of}
binary systems composed of a NS  (of mass $M_{1}$) and a zero-age main sequence (ZAMS)
secondary (of mass $M_{2}$). For the secondary/donor star, we consider three kinds of
chemical compositions (Pop~I with $X=0.70$, $Z=0.02$; Pop~II with
$X=0.75$, $Z=0.001$; and $X=0.7597$, $Z=0.0001$).
We take the ratio of the mixing length to the pressure scale height to be 2.0, and the convective
overshooting parameter to be 0.12. The effective RL radius of the secondary star is given by
the empirical formula proposed by \cite{Eggleton1983},
\begin{equation}
R_\mathrm{L,2}=\dfrac{0.49q^{-2/3}}{0.6q^{-2/3}+\ln(1+q^{-1/3})}a,
\end{equation}
where $q=M_{2}/M_{1}$ is  the mass ratio, and $a$ is the orbital separation.
During the RLO phase, we adopt the mass transfer rate as
$\dot{M}_{2}=-10^{4}[\log(R_{2}/R_\mathrm{L,2})]^{3}\ M_{\odot}\mathrm{yr^{-1}}$,
where $R_{2}$ is the radius of the secondary.

The rate of the orbital angular momentum loss (AML) is composed of three parts:
\begin{equation}
\dot{J}=\dot{J}_\mathrm{GR}+\dot{J}_\mathrm{ML}+\dot{J}_\mathrm{MB}.
\end{equation}
The first term on the right hand side of the above equation is the AML due to
gravitational radiation (GR) \citep{Landau1975}, given by
\begin{equation}
\dot{J}_\mathrm{GR}=-\dfrac{32}{5}\dfrac{G^{7/2}}{c^{5}}\dfrac{M_{1}^{2}M_{2}^{2}(M_{1}+M_{2}^{1/2})}{a^{7/2}},
\end{equation}
where $G$ is the gravitational constant and $c$ is the speed of light.
The second term is due to mass loss. We assume that a fraction $\alpha$ of the transferred
mass is accreted by the NS, and the rest is ejected out of the system from the vicinity of
the NS as isotropic winds, taking away the specific angular momentum of the NS. Then we have
\begin{equation}
\dot{J}_\mathrm{ML}=-(1-\alpha)\dot{M}_{2}(\dfrac{{M}_{2}}{{M}_{1}+{M}_{2}})^{2}a^{2}\omega,
\end{equation}
where $\omega$ is the angular velocity of the binary.
The third AML mechanism is magnetic braking (MB). Here, we adopt the MB formula postulated by  \cite{Verbunt1981} and \cite{Rappaport1983}:
\begin{equation}
\dot{J}_\mathrm{MB}=-3.8\times10^{-30}M_{2}R_{2}^{4}\omega^{3}\ {\rm dyn\,cm}.
\end{equation}
When the convective envelope of the seconday becomes too thin, the MB effect is much reduced.
Following \cite{Podsiadlowski2002}, we add an ad hoc factor
\begin{center}
$\exp(-0.02/q_\mathrm{conv}+1)$ if $q_\mathrm{conv}<0.02$
\end{center}
to Eq.~(5), where $q_\mathrm{conv}$ is the mass fraction of the surface convective envelope.

\section{Results}
In our control model, we consider a binary composed of a NS
of initial mass $M_1=1.4\ M_{\odot}$, and a ZAMS secondary of initial mass
$M_2=1.4\ M_{\odot}$ with Pop~I chemical compositions.
Figure 1 shows the calculated $P_\mathrm{f}-M_\mathrm{WD}$ relation (left panel), and
the $P_\mathrm{i}-P_\mathrm{f}$ relation (right panel), for three different mass transfer
efficiencies, $\alpha = 0$, 0.5 and 1.0.
We calculate the binary evolution down to the lower WD mass limit by gradually reducing
the initial orbital period until the donor star cannot form a WD.
According to our calculations, the $P_\mathrm{f}-M_\mathrm{WD}$
relation extends to $M_\mathrm{WD}\sim (0.14-0.15)\ M_{\odot}$
for $Z=0.02$, $\sim0.16\ M_{\odot}$ for $Z=0.001$, and $\sim0.17\ M_{\odot}$ for $Z=0.0001$.
Additionally there is always a break between the final compact and wide systems,
which is due to the temporary overall contraction of the donor star
when the H-burning shell crosses the hydrogen discontinuity.

Our calculations show that, in converging systems with very late
Case A mass transfer\footnote{Here very late Case A mass transfer
means, at the beginning of RLO, the secondary has evolved near the
end of main sequence but has not developed a He core within it. The
initial conditions of the binary (e.g., the evolutionary status of
the donor at the onset of RLO) and the timescale of  the RLO phase
depend on the efficiency of MB adopted.}, the donor may have time to
develop a He core during the mass transfer, and the
$P_\mathrm{f}-M_{WD}$ relation would be followed. When it leaves the
main sequence and reaches the base of red giant branch, the donor
star consists of a shell-burning layer located between the
degenerate He core and the convective envelope. After the convective
envelope reaches its deepest extent, the H-burning shell has burned
its way out to the discontinuity of higher H abundance left by the
first dredge-up, leading to a slightly lower burning rate and and a
slight decrease in its  radius. Figure 2 shows four evolutionary
tracks of the donor star on the H-R diagram with the initial orbital
periods $P_\mathrm{i}$ lying between 1.057 days and 1.4 days  (left
panel), and its radius evolution (right panel) based on the control
model, where the temporary contraction is emphasized by a circle. In
the cases of relatively long initial orbital period, the donor star
possesses a large convective envelope, so that it still has enough
material in the envelope for burning after the dredge-up. It can
pass through the temporary contraction and expand again to refill
its RL, leaving a relatively massive WD. For relatively short
initial orbital period, the envelope of the donor star has been
stripped to a greater extent. It is unable to refill its RL after
the temporary contraction, or it has already exhausted all the
nuclear fuel and entered the final contraction phase before the
temporary contraction. Hence the break in the
$P_\mathrm{f}-M_\mathrm{WD}$ relation actually reflects whether the
donor star can refill its RL after the first dredge-up.


\subsection{Dependance on the input parameters}

{\noindent\em 1. The Efficiency of Mass Transfer}

There is amounting evidence of nonconservative mass transfer in LMXBs even at
sub-Eddington mass transfer rates \citep[e.g.,][]{Jacobyetal2005,Nice2008,Antoniadis2012}.
Our calculated results with different values of $\alpha= 0$, 0.5 and 1 in Fig.~1 shows
that the $P_\mathrm{f}-M_\mathrm{WD}$ relation and the related discontinuity are
insensitive to $\alpha$. The former has already been noted by many authors.


{\noindent\em 2. The initial NS mass}

Investigations on the formation of  PSR J1614-2230 \citep{Demorest2010}
suggest that it may be born massive \citep{Tauris2011,Lin2011}.
In Fig.~3, we compare the $P_\mathrm{f}-M_\mathrm{WD}$ relations with two different initial
NS mass of 1.4 and $2.0\ M_{\odot}$. Though the increase in the NS mass may influence
the initial parameter space to form BMSPs \citep{Shao2012}, it barely affects the final
$P_\mathrm{f}-M_\mathrm{WD}$ relation \citep{DeVito2010}, so we cannot tell the initial NS mass simply from the current WD mass.

{\noindent\em 3. The initial donor mass}

In Fig.~4, we show the $P_\mathrm{f}-M_\mathrm{WD}$ relations with
initial donor mass of $1.0\ M_{\odot}$, $1.4\ M_{\odot}$, and $2.0\
M_{\odot}$. The left and right panels correspond to $Z=0.02$ and
$Z=0.001$, respectively. In the case of $2.0\ M_{\odot}$ donor star,
systems with massive ($\gtrsim 0.28\ M_{\odot}$) WDs deviate from
the expected relation for less massive donors. The reason is that,
due to the large mass ratio, when the RLO initiates, the thermal
timescale mass transfer is extremely super-Eddington and highly
non-conservative, thus the evolution typically terminates
considerably before this relation is reached
\citep{Podsiadlowski2002,Lin2011}.

{\noindent\em 4. The metallicities}

We compare the $P_\mathrm{f}-M_\mathrm{WD}$ relations for different donor masses
and chemical compositions ($Z=0.02$,
$0.001$, and $0.0001$) in Fig.~5. The reduction of metallicities
leads to smaller stellar radius, shorter nuclear evolutionary timescale, and hence shorter
bifurcation period. Stars with same mass but lower-metallicities form WDs in a more narrow orbit.

\subsection{Comparison with previous works}

The $P_\mathrm{f}-M_\mathrm{WD}$ relation has been investigated
extensively. An empirical fitting formula was first proposed by
\cite{Rappaport1995} for a giant star with \textbf{a} core mass
between $0.15\,M_{\sun}$ and $1.15\,M_{\sun}$ with single star
evolution calculations. \cite{Tauris1999} investigated the evolution
of LMXBs with $P_\mathrm{i} >$ 2 days based on binary evolution
calculations, and presented modified fitting formulae in the range
of $0.18\ M_{\odot} \lesssim M_\mathrm{WD} \lesssim\ 0.45\
M_{\odot}$ for both Pop~I and II stars.  \cite{Ergma1998} showed
that for converging binaries, if the initial orbital periods lie
between the bifurcation period and the so-called boundary orbital
period, the binaries will evolve to short-period BMSPs with a He WD,
the mass of which can be as low as $0.15-0.16\, M_{\sun}$ (for
$Z=0.03$) and  $0.17\, M_{\sun}$ (for $Z=0.003$), while below the
boundary orbital period the binaries will evolve through a period
minimum with a RLO donor and end their evolution as
 ultra-short period LMXBs.
\cite{DeVito2010} and \cite{Lin2011} also systematically calculated the evolution of
LMXBs, and obtained the $P_\mathrm{f}-M_\mathrm{WD}$ relation for Pop~I stars.
More recently, \cite{Smedley2014} showed that the $P_\mathrm{f}-M_\mathrm{WD}$ relation
can be reproduced extending to orbital periods of less than 1 day for a $1M_{\sun}$
giant donor. They also noticed  some discontinuities in the relation around a WD mass
of $0.225\ M_{\odot}$.
In this work we found the lower limit of the WD mass following the $P_\mathrm{f}-M_\mathrm{WD}$
relation to be $\sim (0.14-0.15)\ M_{\odot}$
(for $Z=0.02$), $\sim0.16\ M_{\odot}$ (for $Z=0.001$), and $\sim0.17\ M_{\odot}$ (for $Z=0.0001$),
consistent with previous results.
Our calculations not only confirm the existence of the discontinuities, but also show that
their positions vary with the donor's mass and metallicities.
Generally, for wide systems our results are in accord with
\cite{Tauris1999}, while for the compact systems are \textbf{closer to} \cite{DeVito2010}.
The whole range of the relation agrees well with the fit given by \cite{Lin2011}.

\section{Comparison with observations}
In Fg.~5 we plot 7 BMSPs in which both component masses have been measured.
The mass error bars corresponding to $1\sigma$ uncertainties
\citep[data are taken from][and references therein]{Smedley2014}.
There is broad agreement for both compact and wide systems if a range of chemical abundances
are taken into account.
In the following we discuss two specific cases of BMSPs with low-mass WDs, PSRs J1738+0333 and J0348+0432.

\subsection{PSR J1738+0333}
PSR J1738+0333 is a 5.85 ms pulsar accompanied by a  He WD in a 8.5
hr orbit. It was first discovered with the Parkes 64-m telescope in
a 20-cm survey  in 2001 \citep{Jacoby2005}. The WD companion is
bright enough for optical spectroscopy and photometric study, which
revealed its mass $0.18^{+0.007}_{-0.005}\,M_{\odot}$
\citep{Antoniadis2012}. Combined with the mass ratio $q = 8.1$
inferred from the radial velocities and the precise pulsar timing
ephemeris, the NS mass was inferred to be
$1.47^{+0.07}_{-0.06}\,M_{\odot}$. These estimates imply a highly
nonconservative mass transfer during the LMXB evolution.

For the evolutionary history of this system, \cite{Antoniadis2012}
pointed out that the system is most likely to be the fossil of Case
A RLO, but the WD mass and the orbital eccentricity are both
consistent with Case B RLO, although there is a grey zone for
properties from both cases.


As shown in Figs.~5 and 6 (left panel), PSR J1738+0333 fits well
with the $P_\mathrm{f}-M_\mathrm{WD}$ relation, located between the
cases of $Z=0.001$ and $Z=0.0001$. According to our calculations of
binary evolution with $M_1=1.4M_{\odot}$ and $Z=0.001$, the boundary
between Case A and B RLO is $M_{\rm WD}\sim (0.2-0.3)M_{\odot}$ on
the $P_\mathrm{f}-M_\mathrm{WD}$ relation, for the initial donor
masses $M_2\sim (1.0-2.0)M_{\odot}$. Then the $\sim0.18M_{\odot}$ He
WD companion suggests that PSR J1738+0333 is more likely to have
been recycled through Case A RLO. In the right panel of Fig.~6, we
show three examples of binary evolutionary sequences with the final
WD mass lying between $0.176\ M_{\odot}$ and $0.187\ M_{\odot}$ with
$Z=0.001$. After the RLO the orbit decays due to GR, the rate of
which depends on the WD mass. This may put useful constraints on the
cooling age of the WD.

\subsection{PSR J0348+0432}
PSR J0348+0432 is a a 39 ms pulsar in a 2.46 hr orbit accompanied by
a $0.17\,M_{\odot}$ He WD, discovered with the Green Bank Telescope
\citep{Boyles2013,Lynch2013}. Similar to PSR J1738+0333, the
component masses of the binary were determined by the spectroscopic
and photometric method \citep{Antoniadis2013}. The extreme narrow
orbit and the low mass of the He WD suggest an efficient loss of
orbital angular momentum and an envelope-stripping phase
experienced. These can be achieved either through the common-envelop
(CE) channel or the converging LMXB channel. As pointed out by
\citet{Antoniadis2013}, the formation via a CE and spiral-in phase
is less possible, since the mass transfer in the progenitor binary
would be dynamically stable without forming a CE. In the more
promising converging LMXB channel, according to the estimated
$\sim2$ Gyr cooling age of the WD, the donor star should detach from
its RL at an orbital period of $\sim5$ hr, which was derived from
the GR-induced orbital decay during this time. The main issue is
that converging LMXBs often keep evolving with continuous mass
transfer and evolve to more compact systems. In order to get the
decoupling from the RL at the right values of the orbital period and
the companion mass, a finely tuned termination of the mass-transfer
process at $\sim 5$ hr is required.

As seen in Fig.~5, the $P_\mathrm{f}-M_\mathrm{WD}$ relation at the lower WD mass end
is mainly
affected by the metallicity of the donor. For a given WD mass, the final orbital period decreases
with metallicity. We have calculated the evolution of LMXBs with different
metallicities to form a BMSP with a $\sim0.17\ M_{\odot}$ WD companion, and found that the
orbital period at the decoupling is about 5 hr for $Z=0.0001$, 7 hr for $Z=0.001$, and 14 hr
for $Z=0.02$. The latter two values can be ruled out since they indicate too long a time for orbital
decay after RLO that the lifetime of the pulsar becomes longer
than the Hubble time. In the right panel of Fig.~6, we show the possible formation tracks of
PSR J0348+0432, to demonstrate that it is possible to form PSR J0348+0432-like systems
within the Hubble time if the donor's metallicity was extremely low ($Z=0.0001$). Furthermore,
we mention that the WD cooling time increases
with decreasing metallicity \citep{Serenelli2001,Serenelli2002}, so  the orbital
period at the decoupling could be longer than 5 hr. This may also help to solve the formation
puzzle of PSR J0348+0432.

\section{Conclusions}
The main results in this work are summarized as follows.

1. The $P_\mathrm{f}-M_\mathrm{WD}$ relation for BMSPs with He WD companions can extend to
$\sim (0.14-0.15)\,M_{\odot}$ for $Z=0.02$, $\sim0.16\,M_{\odot}$ for $Z=0.001$,
and $\sim0.17\,M_{\odot}$ for $Z=0.0001$.

2. The $P_\mathrm{f}-M_\mathrm{WD}$ relation is not smooth in reality, but with a discontinuity
at around $0.2\,M_{\odot}$, which separates converging and diverging binaries.

3. BMSPs with low-mass He WD companions in compact binaries can be accounted for
if the progenitor binaries experienced very late Case A  mass transfer. The WD companion of PSR J1738+0333 is
likely to evolve from a Pop~II star. For PSR J0348+0432, to explain its extremely small
orbit, even lower metallicity ($Z=0.0001$) is needed.

\begin{acknowledgements}
We are grateful to an anonymous referee for helpful comments.
This work was supported by the Natural Science Foundation
of China under grant numbers. 11133001 and 11333004, and the Strategic Priority Research Program
of CAS under grant No. XDB09010200.

\end{acknowledgements}


\newpage


\begin{figure}[h,t]
\centerline{\includegraphics[angle=0,width=1.00\textwidth]{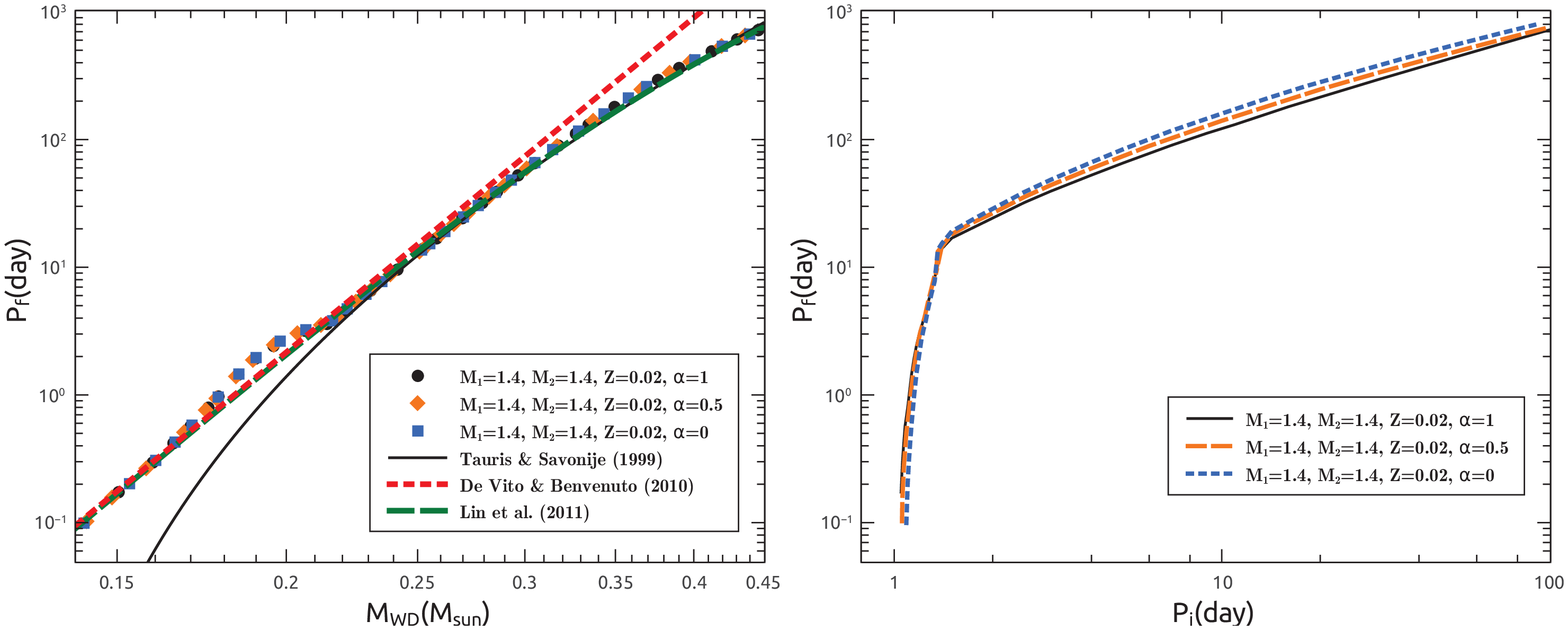}}
\caption{The $P_\mathrm{f}-M_\mathrm{WD}$ relation (left) and $P_\mathrm{i}-P_\mathrm{f}$
relation (right) based on the control model with $\alpha=0$, 0.5 and 1.
The solid, short-dashed, and long-dashed lines in the left panel represent the fitting
formulae given by \cite{Tauris1999}, \cite{DeVito2010} and \cite{Lin2011} for Pop~I stars, respectively. All the masses are in units of Solar mass.
\label{figure1}}
\end{figure}

\newpage
\begin{figure}[h,t]
\centerline{\includegraphics[angle=0,width=1.00\textwidth]{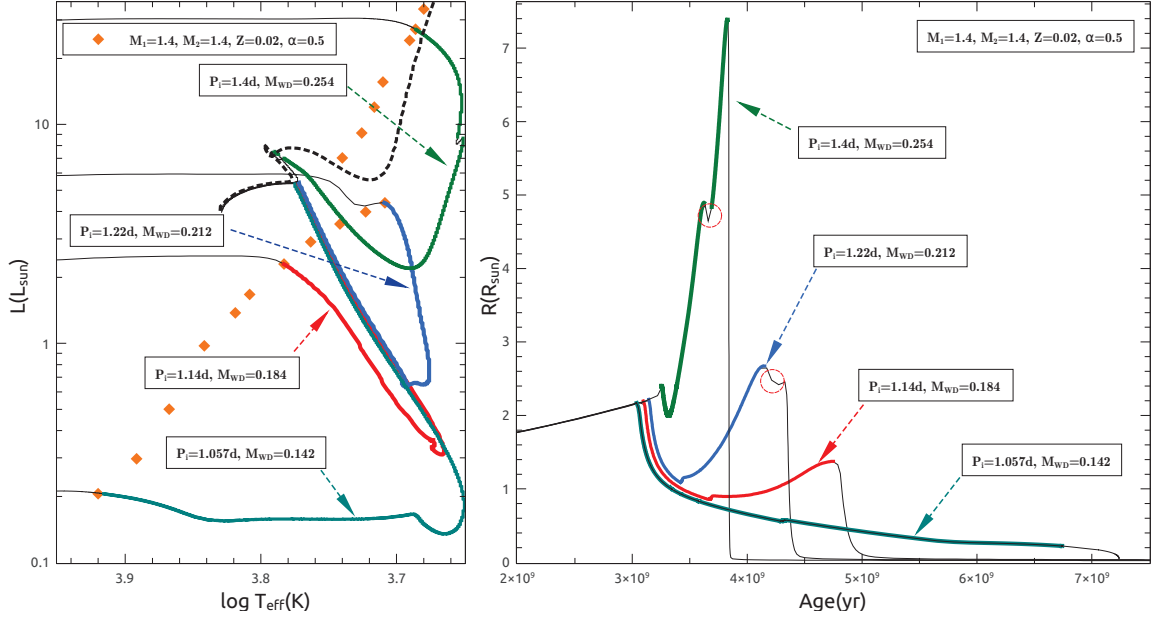}}
\caption{The solid lines in the left panel show the evolutionary
sequences in the control model with different initial orbital
periods, and the thick parts of the lines indicate the RLO phase.
The dashed line represents the evolutionary track of a $1.4\
M_{\odot}$ single star. The diamonds denote the RL detachment of the
donor which gives the $P_\mathrm{f}-M_\mathrm{WD}$ relation. The
lines in the right panel show the corresponding radius evolution of
the donor, and the RLO phase is also denoted by thick lines. The
circles indicate the temporary contraction of the donor due to the
dredge-up process. \label{figure2}}
\end{figure}


\newpage
\begin{figure}[h,t]
\centerline{\includegraphics[angle=0,width=1.00\textwidth]{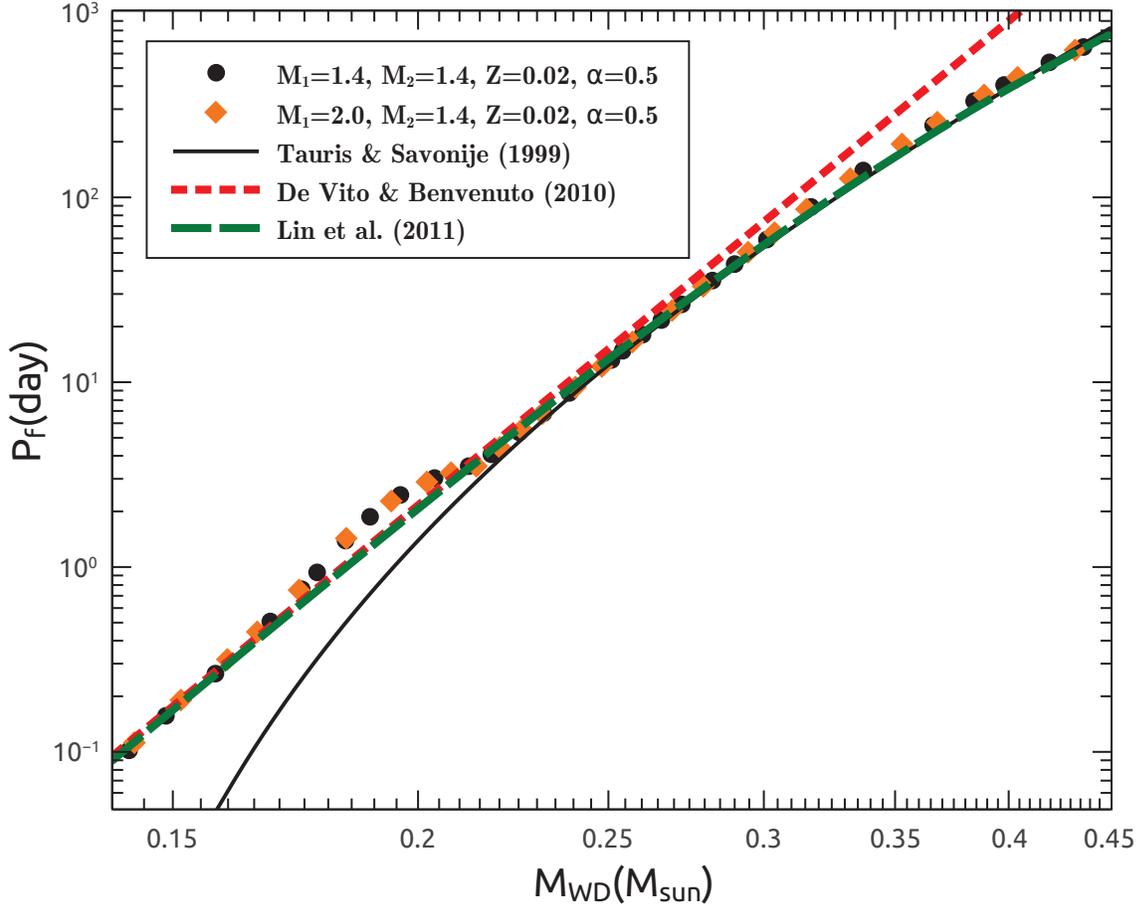}}
\caption{Comparison of the $P_\mathrm{f}-M_\mathrm{WD}$ relations with the
initial NS mass of
$1.4\,M_{\odot}$ and $2.0\,M_{\odot}$ in the control model. The meanings of the
lines are same as in Fig.~1.
\label{figure3}}
\end{figure}

\newpage
\begin{figure}[h,t]
\centerline{\includegraphics[angle=0,width=1.00\textwidth]{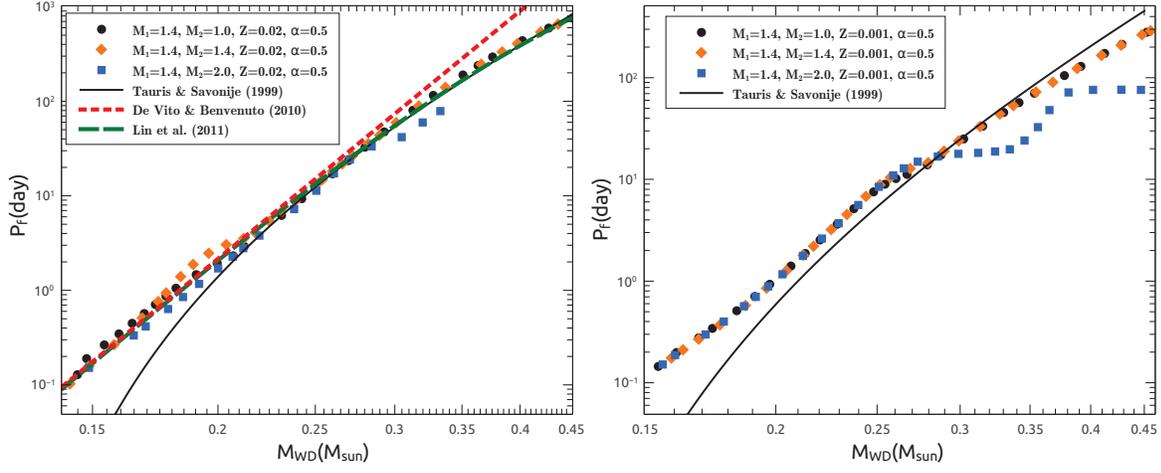}}
\caption{Comparison of the $P_\mathrm{f}-M_\mathrm{WD}$ relations with the initial
donor mass of $1.0\ M_{\odot}$, $1.4\ M_{\odot}$ and $2.0\ M_{\odot}$. The metallicities
are taken to be $Z=0.02$ and 0.001 in the left and right panels, respectively.
The meanings of the lines are same as in Fig.~1.
\label{figure4}}
\end{figure}


\newpage
\begin{figure}[h,t]
\centerline{\includegraphics[angle=0,width=1.00\textwidth]{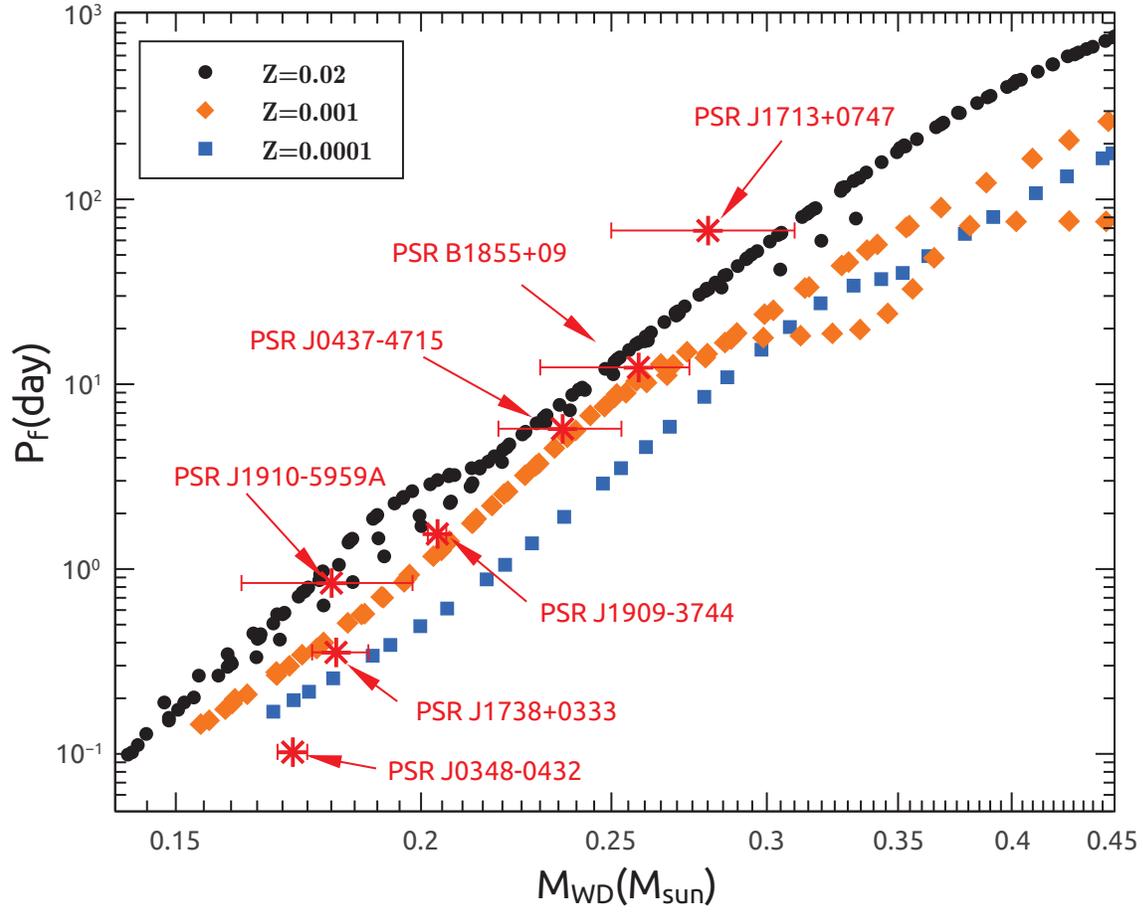}}
\caption{The $P_\mathrm{f}-M_\mathrm{WD}$ relations from all the
calculated results with different metallicities. Also plotted are
BMSPs with both component masses measured. \label{figure5}}
\end{figure}


\newpage
\begin{figure}[h,t]
\centerline{\includegraphics[angle=0,width=1.00\textwidth]{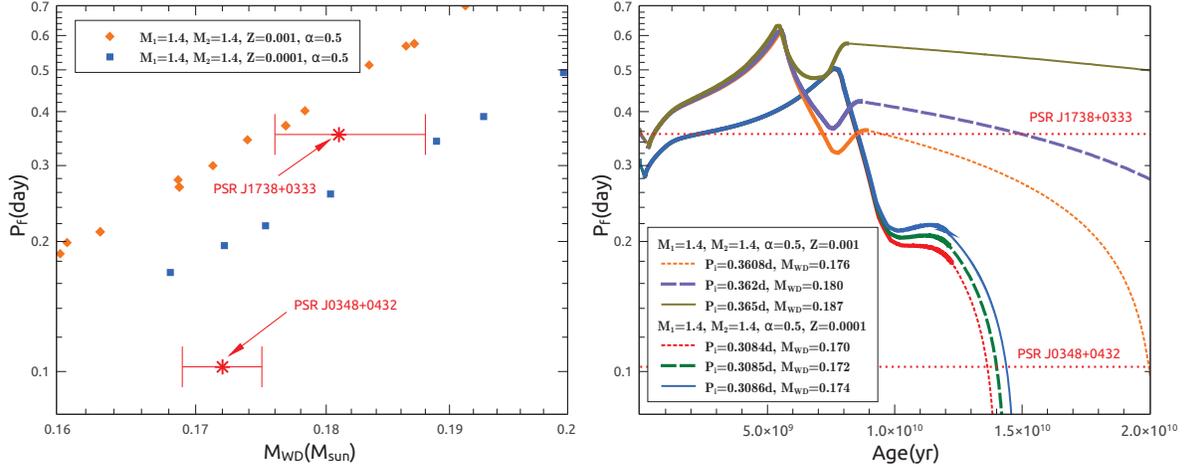}}
\caption{The left panel is part of Fig.~5 for comparison between the
theoretical $P_\mathrm{f}-M_\mathrm{WD}$ relations and the
observations of PSRs J1738+0333 and J0348+0432. The right panel
shows the possible evolutionary tracks for these two pulsars in the
control model, with various initial orbital periods and final WD
masses. The thick solid lines represent the RLO phase, and the
subsequent orbital decay is dominated by GR. The two horizontal
dotted lines represent the current orbital periods of PSRs
J1738+0333 and J0348+0432, respectively. \label{figure6}}
\end{figure}

%


\begin{thebibliography}{28}
\expandafter\ifx\csname natexlab\endcsname\relax\def\natexlab#1{#1}\fi

\bibitem[{{Alpar} et~al.(1982)}]{Alpar1982}
Alpar, A., et al. 1982, Nature, 300, 728

\bibitem[{{Althaus} et~al.(2001)}]{Althaus2001}
Althaus, L. G., Serenelli, A. M., \& Benvenuto, O. G. 2001, MNRAS, 324, 617

\bibitem[{{Antoniadis} et~al.(2012)}]{Antoniadis2012}
Antoniadis, J., van Kerkwijk, M. H., \& Koester, D., et al. 2012, MNRAS, 423, 3316

\bibitem[{{Antoniadis} et~al.(2013)}]{Antoniadis2013}
Antoniadis, J., Freire, P. C. C., Wex, N., Tauris, T. M., \& Lynch, R. S.
et al. 2013, Science, 340, 1233232

\bibitem[{{Archibald} et~al.(2009)}]{Archibald2009}
Archibald, A. M., Stairs, I. H., Ransom, S. M., Kaspi, V. M., \& Kondratiev, V. I.
et al. 2009, Science, 324, 1411

\bibitem[{{Backer} et~al.(1982)}]{Backer1982}
Backer, D. C., Kulkarni, S. R., \& Heiles, C. et al. 1982, Nature, 300, 615

\bibitem[{{Bhattacharya \& van den Heuvel} (1991)}]{Bhattacharya1991}
Bhattacharya D. \& van den Heuvel E. P. J. 1991, Phys. Rep., 203, 1

\bibitem[{{Boyles} et~al.(2013)}]{Boyles2013}
Boyles, J., Lynch, R. S., Ransom, S. M., Stairs, I. H., \& Lorimer, D. R.
et al. 2013, ApJ 763, 80

\bibitem[{{Corongiu} et~al.(2012)}]{Corongiu2012}
Corongiu, A., Burgay, M., \& Possenti, A., et al. 2012, ApJ, 760, 100

\bibitem[{{Deloye} (2008)}]{Deloye2008}
Deloye, C. J. 2008, in AIP Conf. Proc. 983, 40 years of Pulsars: Millisecond
Pulsars, Magnetars and More, ed. C. Bassa, Z. Wang, A. Cumming, \& V. M.
Kaspi (Melville, NY: AIP), 501

\bibitem[{{Demorest} et~al.(2010)}]{Demorest2010}
Demorest, P. B., Pennucci, T., Ransom, S. M., Roberts, M. S., E., \& Hessels, J.
W. T., 2010, Nature, 467, 1081

\bibitem[{{De Vito \& Benvenuto} (2010)}]{DeVito2010}
De Vito, M. A. \& Benvenuto, O. G. 2010, MNRAS, 401, 2552

\bibitem[{{De Vito \& Benvenuto} (2012)}]{DeVito2012}
De Vito, M. A. \& Benvenuto, O. G. 2012, MNRAS, 421, 2206

\bibitem[{{Eggleton} (1971)}]{Eggleton1971}
Eggleton, P. P. 1971, MNRAS, 151, 351

\bibitem[{{Eggleton} (1972)}]{Eggleton1972}
Eggleton, P. P. 1972, MNRAS, 156, 361

\bibitem[{{Eggleton} (1983)}]{Eggleton1983}
Eggleton, P. P. 1983, ApJ, 268, 368

\bibitem[{{Ergma} et~al.(1998)}]{Ergma1998}
Ergma, E., Sarna, M. J., \& Antipova, J. 1998, MNRAS, 300, 352

\bibitem[{{Han} et~al.(2004)}]{Han2004}
Han, Z., \& Podsiadlowski, P. 2004, MNRAS, 350, 1301

\bibitem[{{Jacoby}(2005)}]{Jacoby2005}
Jacoby, B. A. 2005, PhD thesis, California Institute of Technology, California

\bibitem[{{Jacoby} et~al.(2005)}]{Jacobyetal2005}
Jacoby, B. A., Hotan A., Bailes M., Ord S., \& Kulkarni S. R. 2005, ApJ, 629, L113

\bibitem[{{Joss} et~al.(1987)}]{Joss1987}
Joss, P. C., Rappaport, S. A., \& Lewis, W. 1987, ApJ, 319, 180


\bibitem[{{Kilic} et~al.(2010)}]{Kilic2010}
Kilic, M., Brown, W. R., Allende P. C., Kenyon, S. J., \& Panei, J. A. 2010, ApJ, 716, 122

\bibitem[{{Landau \& Lifshitz} (1975)}]{Landau1975}
Landau, L. D., \& Lifshitz, E. M. 1975, in Course of Theoretical Physicsâ
Pergamon International Library of Science, Technology, Engineering and
Social Studies (4th rev. Engl. ed.; Oxford: Pergamon)

\bibitem[{{Lin} et~al.(2011)}]{Lin2011}
Lin, J., Rappaport, S., \& Podsiadlowski, P. et al. 2011, ApJ, 732, 70

\bibitem[{{Lorimer} (2008)}]{Lorimer2008}
Lorimer, D. R. 2008, Living Rev. Relativity, 11, 8

\bibitem[{{Lynch} et~al.(2013)}]{Lynch2013}
Lynch, R. S., Boyles, J., \& Ransom, S. M. et al. 2013, ApJ, 763, 81

\bibitem[{{Ma \& Li} (2009)}]{Ma2009}
Ma, B. \& Li, X.-D. 2009, ApJ, 691, 1611

\bibitem[{{Marsh, Nelemans \& Steeghs} (2004)}]{Marsh2004}
Marsh, T. R., Nelemans, G., \& Steeghs, D. 2004, MNRAS, 350, 113

\bibitem[{{Marsh et al.} (2008)}]{Nice2008}
Nice, D. J., Stairs, I. H., \& Kasian, L. E. 2008, in 40 Years of Pulsars:
Millisecond Pulsars, Magnetars and More. eds. Bassa C.,Wang Z., Cumming
A., Kaspi V. M., AIP Conf. Ser.  (Am. Inst. Phys., New York), Vol. 983,
p. 453

\bibitem[{{Panei} et~al.(2007)}]{Panei2007}
Panei, J. A., Althaus, L. G., Chen, X., \& Han, Z. 2007, MNRAS, 382, 779

\bibitem[{{Papitto} et~al.(2013)}]{Papitto2013}
Papitto, A., Ferrigno, C., \& Bozzo, E., et al. 2013, Natur, 501, 517

\bibitem[{{Patruno \& Watts} (2012)}]{Patruno2012}
Patruno, A. \& Watts, A. L. 2012, arXiv:1206.2727

\bibitem[{{Podsiadlowski} et~al.(2002)}]{Podsiadlowski2002}
Podsiadlowski, P., Rappaport, S., \& Pfahl, E. D. 2002, ApJ, 565, 1107

\bibitem[{{Pols} et~al.(1995)}]{Pols1995}
Pols, O. R., Tout, C. A., Eggleton, P. P., \& Han, Z. 1995, MNRAS, 274, 964

\bibitem[{{Pylyser \& Savonije} (1988)}]{Pylyser1988}
Pylyser, E., \& Savonije, G. J. 1988, A\&A, 191, 57

\bibitem[{{Pylyser \& Savonije} (1989)}]{Pylyser1989}
Pylyser, E., \& Savonije, G. J. 1989, A\&A, 208, 52

\bibitem[{{Radhakrishnan \& Srinivasan} (1982)}]{Radhakrishnan1982}
Radhakrishnan, V. \& Srinivasan, G. 1982, Current Science, 51, 1096

\bibitem[{{Rappaport} et~al.(1983)}]{Rappaport1983}
Rappaport, S., Verbunt, F., \& Joss, P. C. 1983, ApJ, 275, 713

\bibitem[{{Rappaport} et~al.(1995)}]{Rappaport1995}
Rappaport, S., Podsiadlowski, Ph., Joss, P. C., Di Stefano, R., \& Han, Z. 1995, MNRAS, 273, 731

\bibitem[{{Refsdal \& Weigert} (1971)}]{Refsdal1971}
Refsdal, S., \& Weigert, A. 1971, A\&A, 13, 367

\bibitem[{{Serenelli} et~al.(2001)}]{Serenelli2001}
Serenelli, A. M. Althaus, L. G.; Rohrmann, R. D., \& Benvenuto, O. G. 2001, MNRAS, 325, 607

\bibitem[{{Serenelli} et~al.(2002)}]{Serenelli2002}
Serenelli, A. M., Althaus, L. G., Rohrmann, R. D., \& Benvenuto, O. G. 2002, MNRAS, 337, 1091

\bibitem[{{Shao \& Li} (2012)}]{Shao2012}
Shao, Y. \& Li, X.-D. 2012, ApJ, 756, 85

\bibitem[{{Smedley} et~al.(2014)}]{Smedley2014}
Smedley, S. L., Tout, C. A., Ferrario, L., \& Wickramasinghe, D. T. 2014, MNRAS, 437, 2217

\bibitem[{{Splaver} et~al.(2005)}]{Splaver2005}
Splaver, E. M., Nice, D. J., Stairs, I. H., Lommen, A. N., \& Backer D. C. 2005, ApJ, 620, 405

\bibitem[{{Tauris et al.} (2011)}]{Tauris2011}
Tauris ,T. M., Langer, N., \& Kramer, M. 2011, MNRAS, 416, 2130

\bibitem[{{Tauris \& Savonije}(1999)}]{Tauris1999}
Tauris, T. M., \& Savonije, G. J. 1999, A\&A, 350, 928

\bibitem[{{Tauris \& van den Heuvel} (2006)}]{Tauris2006}
Tauris T. M. \& van den Heuvel E. P. J. 2006, in Compact
stellar X-ray sources. ed. W. Lewin \& M. van der Klis
(Cambridge: Cambridge Univ. Press), 623

\bibitem[{{van Kerkwijk} et~al.(2005)}]{van Kerkwijk2005}
van Kerkwijk, M., Bassa, C. G., Jacoby, B. A., \& Jonker, P. G.
2005, in Binary Radio Pulsars, F.
Rasio, I. H. Stairs, eds., vol. 328 of Astronomical Society of the Pacific Conference Series
(San Francisco, 2005), p. 357.

\bibitem[{{van der Sluys} et~al.(2005a)}]{van der Sluys2005a}
van der Sluys, M. V., Verbunt, F., \& Pols, O. R. 2005a, A\&A, 431, 647

\bibitem[{{van der Sluys} et~al.(2005b)}]{van der Sluys2005b}
van der Sluys, M. V., Verbunt, F., \& Pols, O. R. 2005b, A\&A, 440, 973

\bibitem[{{van Straten} et~al.(2001)}]{van Straten2001}
van Straten, W., Bailes, M., \& Britton, M. et al. 2001, Nature, 412, 158

\bibitem[{{Verbunt \& Zwaan} (1981)}]{Verbunt1981}
Verbunt, F., \& Zwaan, C. 1981, A\&A, 100, L7

\bibitem[{{Webbink} et~al.(1983)}]{Webbink1983}
Webbink, R. F., Rappaport, S. A., \& Savonije, G. J. 1983, ApJ, 270, 678

\bibitem[{{Zhang} et~al.(2011)}]{Zhang2011}
Zhang, C. M., Wang, J., \& Zhao, Y.H. et al. 2011, A\&A, 527, 83

\end{thebibliography}
\end{document}